\documentclass[12pt]{article} 
\usepackage[tablesfirst,notablist,nomarkers]{endfloat} 

\usepackage{ol2}
\usepackage[draft]{hyperref}
\usepackage{amsmath}
\usepackage[applemac]{inputenc}
\usepackage{color}
\usepackage{bm}
\usepackage{graphicx}

\begin{document}


\title{Conical emission from laser filaments 

and higher-order Kerr effect in air}


\author{P.  B\'ejot and J. Kasparian$^{*}$}
\address{
Universit\'e de Gen\`eve, GAP-Biophotonics, 20 rue
de l'Ecole de M\'edecine, 1211 Geneva 4, Switzerland
 \\
$^*$Corresponding author: jerome.kasparian@unige.ch
}
\begin{abstract}
We numerically investigate the conical emission (CE) from ultrashort laser filaments, both considering and disregarding the higher-order Kerr effect (HOKE). While the consideration of HOKE has almost no influence on the predicted CE from collimated beams, differences arise for tightly focused beams. This difference is attributed to the different relative contributions of the non-linear focus and of the modulational instability over the whole filament length.
\end{abstract}

\ocis{320.2250, 190.2620, 190.3270, 190.5940, 190.7110, 320.7110}


The filamentation of ultrashort laser pulses attracts much interest due to its spectacular potential applications \cite{KasparianW08} such as remote sensing
\cite{KasparianRMYSWBFAMSWW03}, lightning control \cite{KasparianAAMMPRSSYMSWW08}, water condensation \cite{RohwetterKSHLNPQSSHWW09}, or the generation of THz radiation \cite{TzortMPAPFMMGBE2002}.
A prominent property of filaments is conical emission (CE) \cite{NibberingCGPFSM96,KosarevaKBCC97}, which denotes the spatio-spectral distribution of the supercontinuum generated in the filaments, with a well-defined relationship between the wavelength and the emission angle. It provides an angle-resolved spectrum reflecting the complex dynamics of femtosecond pulses propagating in nonlinear dispersive media \cite{FaccioTMBBLDDM05}.
CE is observed on both sides of the fundamental wavelength \cite{ThebergeCRMD08}. This spectral symmetry is  understood by considering either modulational instability
\cite{LutheMNW1994,B'ejoKHLF2011} or the X-wave propagation regime \cite{FaccioTMBBLDDM05}.

Filamentation is generally described as a dynamic balance between Kerr self-focusing and defocusing by the plasma generated at the non-linear focus \cite{ChinHLLTABKKS05,BergeSNKW07,CouaironM07,KasparianW08}. 
The  temporally asymmetric distribution of the plasma along the pulse local time is expected to generate a spectral asymmetry of CE in the region of the filament onset (or beam collapse). \cite{BergeSNKW07,CouaironM07,FaccioALTCPT08,KosarevaKBCC97}. 

However, a controversy is currently running about the possible contribution of higher-order Kerr effect (HOKE) \cite{LoriotHFL09} to filamentation in gases  \cite{LoriotBHFLHKW09,B'ejoHFL2011,PolynKWM2010,KosarDPWHYRMC2011}. Since their consideration leads to expect 10 to 100 times lower plasma densities in filaments \cite{LoriotBHFLHKW09}, and CE is both a key property of filaments and expectedly strongly connected to ionization, it is important to understand how the HOKE would impact CE.


In this Letter, we numerically investigate both the impact of beam focusing on CE and the build-up of the CE along the filament length.
We find that both models reproduce CE on both sides of the fundamental wavelength for collimated beams, but that considering the HOKE is necessary to reproduce the observed CE  from focused beams on the high-frequency side of the spectrum \cite{FacciALDCPT2008}.


We simulated the propagation of a 5 mJ, 45~fs pulse (13 critical powers $P_{cr}$) centered at 800 nm, launched collimated in air with a radius of 3.8 mm at $1/e^2$. The code is based on the unidirectional pulse propagation equation (UPPE) \cite{KolesikMM02}, in which we used the published values for the HOKE \cite{LoriotHFL09}. The model, described in detail in \cite{B'ejoHFL2011} in the case of argon, was completed by a detailed {weak-field quantum modeling of the Raman-induced} rotational effects in the air \cite{BarteWLKM2002}.
The far-field CE pattern is computed as {Fourier (in the temporal domain) and Hankel (in the spatial domain)} transforms of the electric field.

\begin{figure}[t]
   \begin{center}
     \includegraphics[keepaspectratio,width=12.5cm]{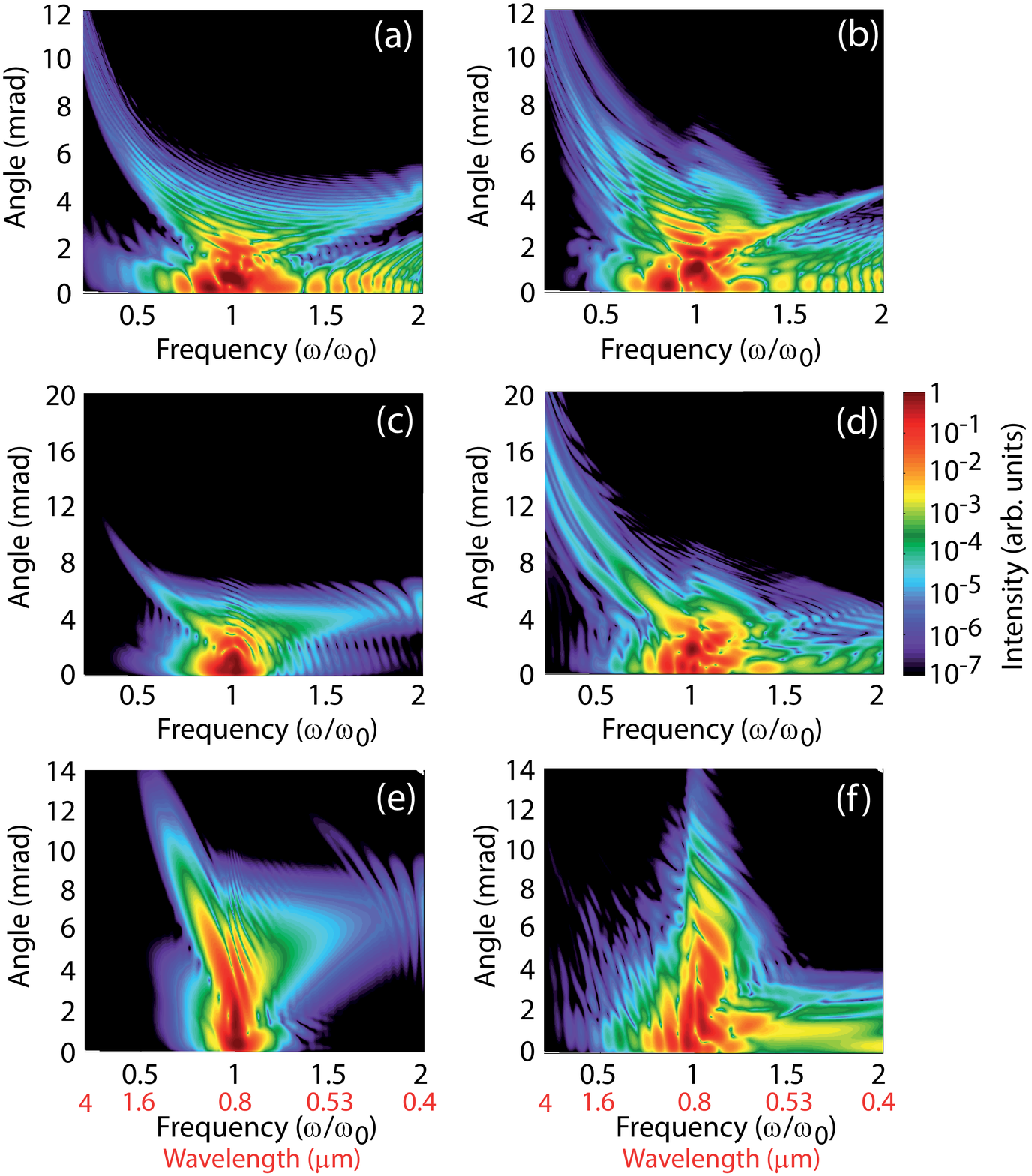}
   \end{center}
   \caption{(Color online) Far-field angularly-resolved spectrum  of 1 mJ, 30 fs pulses (4 $P_{cr}$) centered at 800 nm, with a radius of 3~mm at $1/e^2$. (a,b) $f$ = 4 m, f/\# = 1333; (c,d) $f$ = 2 m, f/\# = 667; (e,f) $f$ = 1 m, f/\# = 333, in the case of the full model considering the HOKE (a,c,e) and of the truncated model disregarding the HOKE (b,d,f)}
   \label{CE_foc}
\end{figure}

Both the behavior of the conical emission and the impact of the HOKE consideration strongly depend on external focusing. For a collimated beam, both models reproduce the CE  from a collimated beam at the experimentally measured angles on both the visible \cite{MaioliSLSBKW09} and the infrared \cite{ThebergeCRMD08} sides of the spectrum. However, predictions diverge when the focusing gets tighter. The truncated model predicts the disappearance of the CE, first on the {visible} side (Figure \ref{CE_foc}d), then on the infrared side (Figure \ref{CE_foc}f) of the fundamental wavelength.
Conversely, the full model predicts CE on both sides of the spectrum, regardless of the numerical aperture. Focusing only offsets the CE angle by an amount comparable to the numerical aperture imposed to the beam, and broadens the angular {distribution} for each spectral component (Figure \ref{CE_foc}a,c,e). 
Let us note that  CE was indeed observed in the visible side of the spectrum in beams with comparable $f$-numbers by Faccio et al. \cite{FaccioALTCPT08}. Such behavior is only reproduced by the full model, although experimental conditions  differ slightly from those of our calculations, preventing a direct comparison.

\begin{figure}[t]
   \begin{center}
      \includegraphics[keepaspectratio,width=12.5cm]{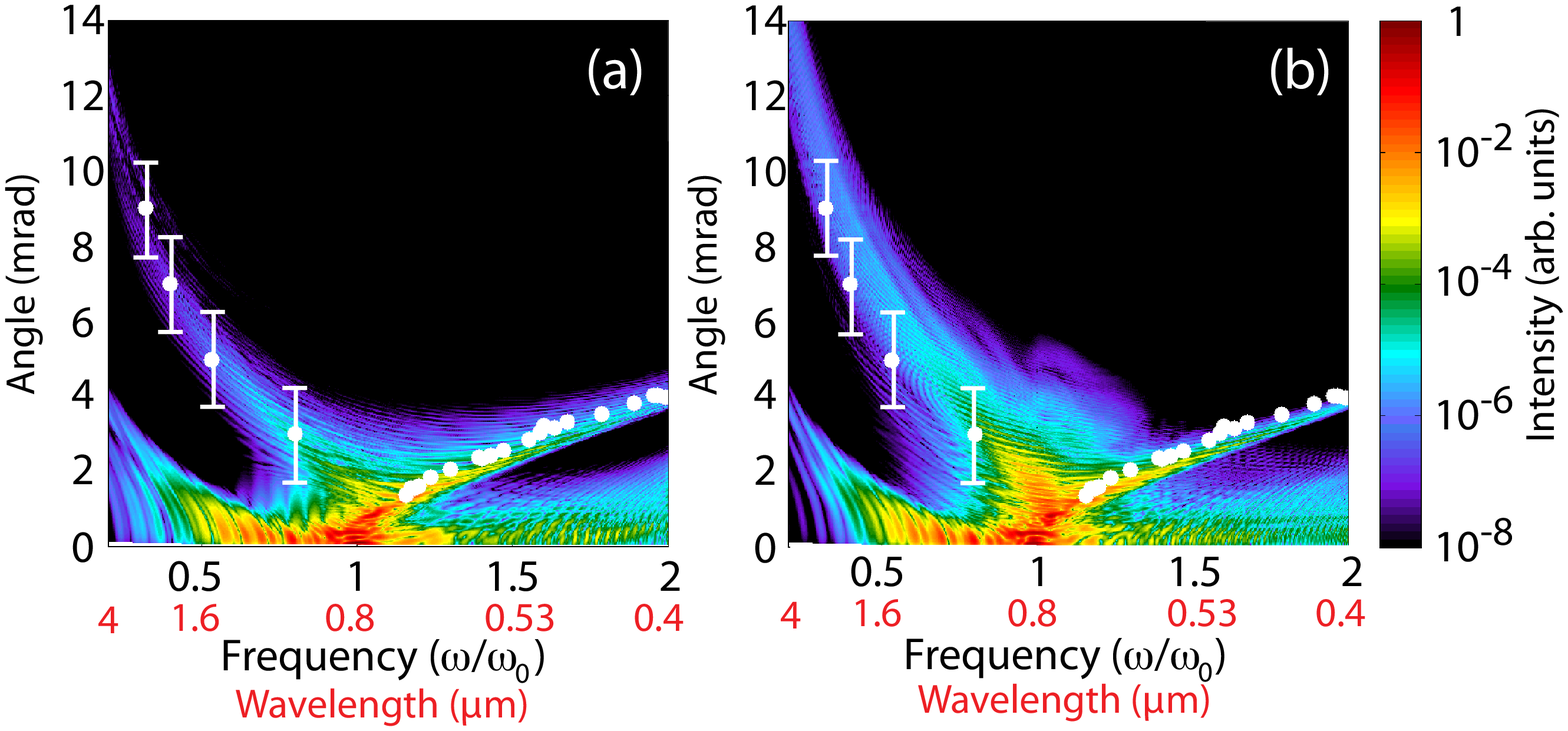}
   \end{center}
   \caption{(Color online) Far-field angularly-resolved spectrum  of 5 mJ, 45~fs pulses centered at 800 nm, after 24 m of propagation {in air}, including 16 m of filamentation. (a) Full model considering the HOKE; (b) Truncated model disregarding the HOKE. White dots display the experimental data of Th\'eberge et al. \cite{ThebergeCRMD08} and Maioli et al. \cite{MaioliSLSBKW09} on the IR and visible sides of the spectrum, respectively. Media 1 displays the build-up of these patterns over 8--14~m propagation range.}
   \label{Figure1_CE_exists}
\end{figure}

\begin{figure}[t]
   \begin{center}
      \includegraphics[keepaspectratio,width=12.5cm]{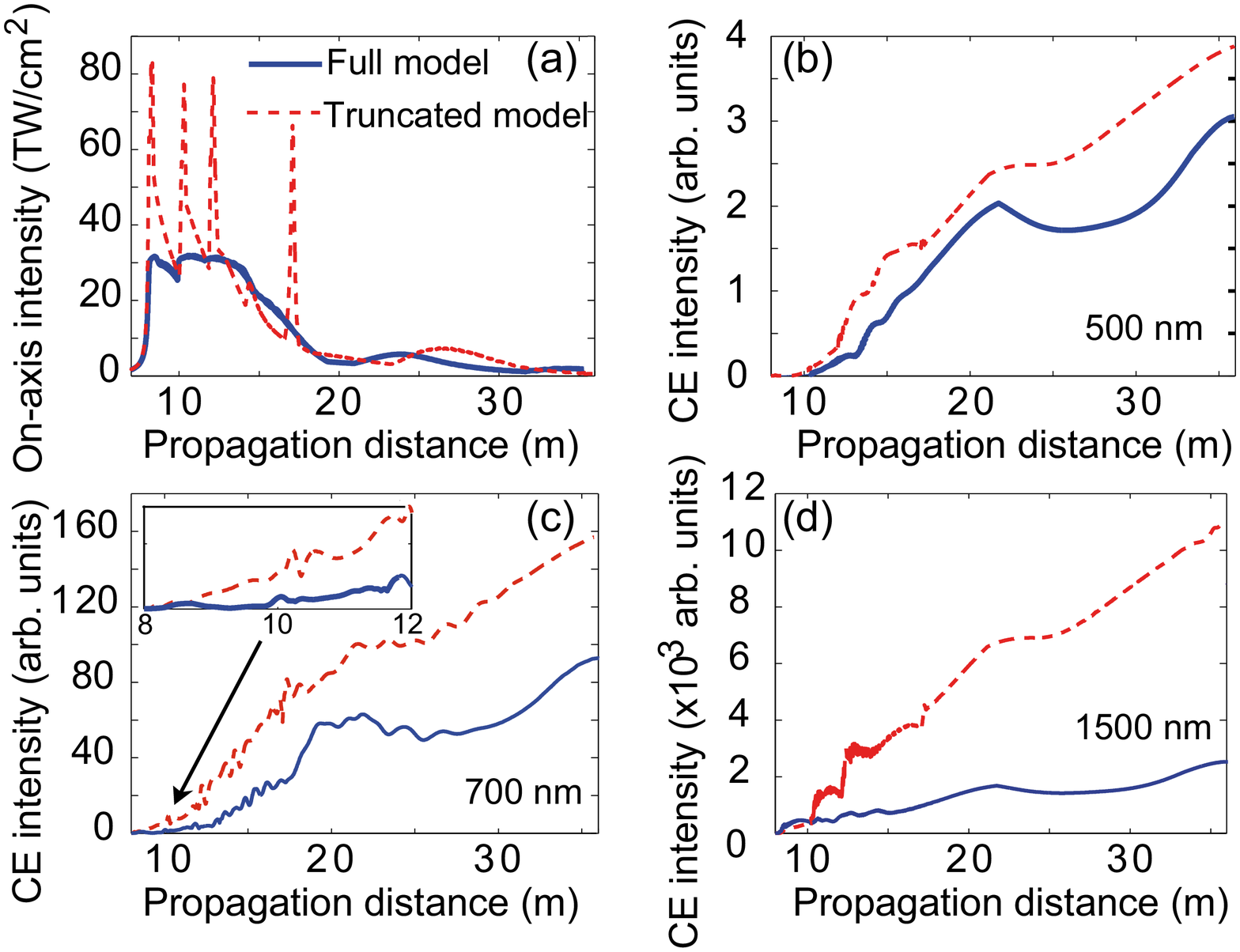}
   \end{center}
   \caption{(Color online) Build-up of the conical emission from a 5 mJ, 45~fs pulse centered at 800 nm.
   (a) Total on-axis intensity. 
   (b-d) Intensity of the CE lobe at (b) 500 nm;  (c) 700 nm and (d) 1500 nm as a function of propagation distance, as predicted by the full model considering the HOKE and the truncated model disregarding the HOKE. Note that panels b-d share a common intensity scale.}
   \label{CE_build_up}
\end{figure}

The similarity of the conical emission patterns predicted by both models from a collimated beam (Figure \ref{Figure1_CE_exists}) apparently contradicts the recent finding that considering the HOKE in models suppresses the CE from the filament onset region \cite{KosarDPWHYRMC2011}. This paradox can be resolved by considering the build-up of CE over the propagation distance (Media 1 of Figure \ref{Figure1_CE_exists} and Figure \ref{CE_build_up}). Note that, in the truncated model outcome, the intensity spikes are not numerical artifacts. They stem from refocusing of the the trailing part of the pulse  during the propagation \cite{GaardeC2009}. 
Although both models predict the same filamentation onset distance ($z\approx9$~m, Figure \ref{CE_build_up}a), the truncated one yields a slightly earlier rise of the CE, which is more pronounced on the visible side of the spectrum (Figure \ref{CE_build_up}b,c). 
This is especially the case at wavelengths close to the fundamental one, where the build-up of the CE appears offset by up to 1 m between the two models (See inset of Figure \ref{CE_build_up}c).
The offset in the CE build-up predicted by the two models at the filament onset may in principle be used to distinguish between their validity  \cite{KosarDPWHYRMC2011}. However,  this offset has the same order of magnitude than the mismatch displayed the filament onset location observed experimentally and predicted by the full model on one side, and its location predicted by the truncated model on the other side (See Figure 1a,b of \cite{KosarDPWHYRMC2011}). We therfore expect that such test is hardly conclusive in practice.

The differences observed between the two models including and disregarding the HOKE can be understood by considering that two processes contribute to the CE, namely the modulational instability (MI) all along the filament length, and refraction of the continuum by the refractive index gradient induced by the transverse plasma density profile.
The gain of the MI contribution to CE writes \cite{B'ejoKHLF2011}:
\begin{equation}
K=\sqrt{D \left(2k_0I_0n_{2,\textrm{eff}}-D \right) + {k^{(1)}}^2 \omega^2 I_0^2 n_{2,\textrm{eff}}^2}
\end{equation}
where $D=\frac{k_{\perp}^2}{2k_0}-\frac{k^{(2)}\omega^2}{2}$ is the spatio-spectral dispersion operator,  $\omega$ is the frequency detuning relative to the fundamental frequency, $k_0$ (taken at $\omega_0$) and $k_{\perp}$ the total and transverse wave vectors, $k^{(1)}$ and $k^{(2)}$ the first- and second-order dispersion term, $I_0$ the incident intensity and $n_{2,\textrm{eff}}$ the effective non-linear refractive index. For $n_{2,\textrm{eff}}<0$, the gain cancels and the supercontinuum emission is confined on the beam axis \cite{B'ejoKHLF2011}. 
Conversely, since the electron density accumulates over the pulse duration, its contribution intrinsically bears a strong temporal asymmetry, which converts into a spectral one, because dispersion and self-phase modulation red-shift the front of the pulse and blue-shift its trail. Furthermore, the plasma contribution can be considered as negligible in the framework of the full model, where the plasma density is insufficient to significantly influence the pulse propagation \cite{B'ejoHFL2011}.

Compared to the full model, the truncated one predicts a nonlinear refractive index inversion at a higher intensity \cite{LoriotBHFLHKW09}, resulting in a higher MI gain, hence in an earlier growth of CE at the filament onset, followed by a larger intensity. Furthermore, the high-intensity peak predicted at $z\approx9$~m by the truncated model (Figure \ref{CE_build_up}a) will result in high plasma density at the filament onset, offering a supplementary boost to the early CE emission on the visible side of the spectrum. 

Tighter focusing results in higher plasma densities, especially in the framework of the truncated model. This  plasma density accumulating during the pulse results in an increasingly negative contribution to $n_{2,\textrm{eff}}$, ultimately inverting it to negative values preventing CE. Focusing tighter will yield to an earlier inversion of $n_{2,\textrm{eff}}$, extinguish CE from MI first on the visible side of the fundamental, and then on the infrared side, as displayed in Figure \ref{CE_foc}d,f.

In conclusion, we have  investigated the influence of beam focusing on conical emission as well as the dynamics of its  build-up along the filamentation of ultrashort laser pulses. 
CE is emitted from the whole filament length through MI \cite{B'ejoKHLF2011}. For collimated beams, which yield long filaments, the contribution of the non-linear focus region is moderate and the consideration of the HOKE has no practical impact on the predicted CE. 
For tightly focused beams, the high plasma density predicted if the HOKE are disregarded would lead to expect the extinction of CE.
Our work therefore illustrates the need to consider CE emission from the whole filament length. It also suggests the behavior of CE at tight focusing as a successful experimental test of the need to consider the HOKE to adequately describe CE from laser filaments.

Acknowledgements. We acknowledge fruitful discussion with O. Kosareva and S. Tzortzakis. This work was supported by the 
Swiss NSF (contract 200021-125315).

\bibliographystyle{unsrt}


\end{document}